\def\be{\begin{equation}}
\def\ee{\end{equation}}
\def\bea{\begin{eqnarray}}
\def\eea{\end{eqnarray}}

\documentclass[aps,prb,twocolumn,amsmath,amssymb,showpacs]{revtex4}
\usepackage{epsfig,graphicx}
\usepackage{bm}

\begin{document}


\title{Analysis of a tuneable coupler for superconducting phase qubits}

\author{Ricardo A. Pinto}
\affiliation{Department of Electrical Engineering, University of
California, Riverside, CA 92521, USA}
\author{Alexander N. Korotkov\footnote{Corresponding author; korotkov@ee.ucr.edu}}
\affiliation{Department of Electrical Engineering, University of
California, Riverside, CA 92521, USA}
\author{Michael R. Geller}
\affiliation{Department of Physics and Astronomy, University of
Georgia, Athens, Georgia 30602, USA}
\author{Vitaly S. Shumeiko}
\affiliation{Department of Microtechnology and Nanoscience, MC2,
Chalmers University of Technology, SE-41296 Gothenburg, Sweden}
\author{John M. Martinis}
\affiliation{Department of Physics, University of California, Santa
Barbara, California 93106, USA}

\date{\today}


\begin{abstract}
This paper presents a theoretical analysis of the recently realized
tuneable coupler for superconducting phase qubits (R. C. Bialczak et
al., Ref.\ \protect\onlinecite{Bialczak}). The coupling can be
turned off by compensating a negative mutual inductance with a
tuneable Josephson inductance. The main coupling in this system is
of the $XX$ type and can be zeroed exactly, while there is also a
small undesired contribution of the $ZZ$ type. We calculate both
couplings as functions of the tuning parameter (bias current) and
focus on the residual coupling in the OFF regime. In particular, we
show that for typical experimental parameters the coupling OFF/ON
ratio is few times $10^{-3}$, and it may be zeroed by proper choice
of parameters. The remaining errors due to physical presence of the
coupler are on the order of $10^{-6}$.


\end{abstract}


\maketitle


\section{Introduction}

    Superconducting qubits \cite{scqubits} are potential building blocks of
a quantum computer. Among their advantages in comparison with other
qubit types are an efficient control with voltage/current/microwave
pulses and use of a well-developed technology suitable for large
scale integration.
 Recent demonstrations of simple quantum
algorithms \cite{Yale-Nature-09,Yamamoto-10} and three-qubit
entanglement \cite{3qubit-UCSB,3qubit-Yale} with superconducting
qubits are important steps towards a practical quantum computation.

    In the standard idea of a gate-based quantum computation \cite{N-C}
it is important that the qubits are decoupled from each other for
most of the time, and the coupling of a desired type between two (or
three) qubits is switched on only for a quantum gate operation,
while it is switched off again after that. Since the superconducting
qubits cannot be physically moved in space, such coupling/decoupling
should be realized by changing control parameters of a circuit. The
simplest idea is to tune the qubits in resonance with each other for
efficient coupling and move them out of resonance for decoupling
(see., e.g., Refs.\
\onlinecite{Yale-Nature-09,Yamamoto-10,3qubit-UCSB,3qubit-Yale}).
However, this requires avoiding unwanted resonances, and with
increasing number of qubits may lead to the problem of ``spectral
crowding''. Even more important limitation of this approach is the
following. Because the effective coupling strength when the qubits
are detuned by an energy $\Delta E$ is of order $g^2/\Delta E$,
where $g$ is the tuned value, the ratio of the ``switched off" and
``switched on'' coupling strengths is ${\rm OFF/ON} \simeq g/\Delta
E$. This ``OFF/ON ratio'' characterizes a coupler's ability to
successfully turn on and off the coupling between qubits. $\Delta
E/h$ is limited to about a GHz in current superconducting qubit
devices. To make the ratio small then requires $g$ to be small,
which makes gate operations slow. For example, to realize an OFF/ON
ratio of $10^{-3}$ when $\Delta E/h=1 \, {\rm GHz}$ would require
$g$ to be $1 \, {\rm MHz}$, which is unacceptably small.

 A different idea is to introduce an extra element between
the qubits: an adjustable coupler, which can turn the coupling on
and off. It is a much better approach from the architecture point of
view since it allows easier design of complex quantum circuits. This
has motivated several experimental
\cite{Hime,Niskanen,Allman,Harris,Ploeg,Bialczak} and theoretical
\cite{Blais-tunable,Averin-03,Filippov-03,Lantz2004,Plourde-th,Wallquist2005,Niskanen-06,Hutter-06,Geller07,Nori,Peropadre}
studies of adjustable couplers for superconducting qubits.

    In this paper we theoretically analyze operation of a recently realized
\cite{Bialczak} tuneable coupler for superconducting phase qubits,
which demonstrated current-controlled tuning of the $XX$ type
coupling from 0 to 100 MHz. In the next Section we discuss the
Hamiltonian of the analyzed Josephson circuit and our definition of
the two-qubit coupling frequencies $\Omega_{XX}$ and $\Omega_{ZZ}$,
corresponding to the $XX$ and $ZZ$ types of interaction (for our
system typically $|\Omega_{XX}|\gg |\Omega_{ZZ}|$, so the main
coupling is of the $XX$ type). In Sec.\ III we find $\Omega_{XX}$
and $\Omega_{ZZ}$ in a simple semiclassical way, while in Sec.\ IV
similar results are obtained in the lowest-order quantum analysis.
We show that both $\Omega_{XX}$ and $\Omega_{ZZ}$ can cross zero as
functions of the control parameter (bias current), but typically not
simultaneously, thus leading to a non-vanishing residual coupling,
which is discussed in Sec.\ V. In particular, we show that typical
OFF/ON ratio for the coupler is few times $10^{-3}$; however, a
minor modification of the experimental circuit \cite{Bialczak}
(addition of a small coupling capacitance) can zero the residual
coupling, thus zeroing the OFF/ON ratio. Actually, this does not
mean complete decoupling, because we rely on the two-qubit
description of a more complicated circuit. The remaining coupling
effects are also discussed in Sec.\ V and are shown to lead to
errors on the order of $10^{-6}$. Section VI presents numerical
results of the quantum analysis; for typical experimental parameters
they are close to the analytical results. Section VII is the
Conclusion.
    In Appendix we discuss the position and momentum matrix elements
for an oscillator with weak cubic nonlinearity, and derive improved
analytics for $\Omega_{XX}$ and $\Omega_{ZZ}$.

\section{System and Hamiltonian}
\label{sec-system}
  \vspace{-0.1cm}

\begin{figure}[tb]
\vspace{-0.3cm}
  \centering
\includegraphics[width=8.0cm]{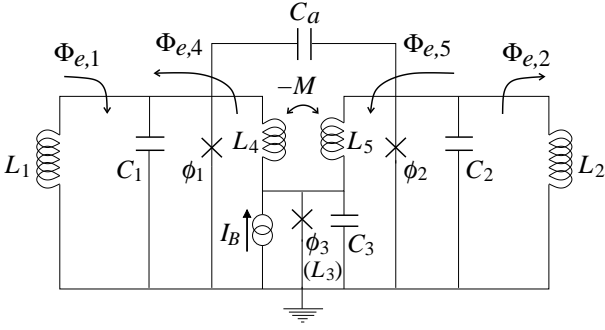}
 \vspace{-0.2cm}
  \caption{The analyzed scheme of two-qubit coupling, which is controlled
by the bias current $I_B$ of the coupling Josephson junction.
Indices 1 and 2 refer to the two qubits, and index 3 refers to the
coupling junction. Coupling inductors $L_4$ and $L_5$ have a
negative mutual inductance $-M$. The current $I_B$ controls the
Josephson inductance $L_3$ of the coupling junction, which
effectively adds to $-M$.
 An additional small coupling is via
capacitance $C_a$. }
  \label{Schematic}
\end{figure}

    Let us consider the system \cite{Bialczak} shown in Fig.\
\ref{Schematic}, which consists of two flux-biased phase qubits
\cite{Cooper-04} characterized by capacitances $C_{1}$ and $C_2$,
inductances $L_{1}$ and $L_2$, and Josephson energies $E_{J1}$ and
$E_{J2}$ of the junctions (in Fig.\ \ref{Schematic} the
superconducting phases across these Josephson junctions are denoted
as $\phi_1$ and $\phi_2$).
 The qubits are coupled via an additional Josephson
junction (characterized by $C_3$ and $E_{J3}$) with an adjustable
bias current $I_B$; the qubits are connected to this junction via
inductances $L_{4}$ and $L_5$, which have a negative mutual
inductance $-M$ ($M>0$). We also introduce the qubit coupling via a
very small capacitance $C_a$, which was not implemented in the
experiment, \cite{Bialczak} but may be important in future
experiments for turning the coupling off more precisely.

    The general idea of this scheme \cite{Bialczak} is that the
Josephson inductance $L_3$ of the middle junction is essentially in
series with the mutual inductance $-M$, and therefore (in absence of
$C_a$) the coupling is expected to be (crudely) proportional to
$L_3-M$. Then varying $L_3$ by varying the bias current $I_B$, it is
possible to adjust the qubit-qubit coupling strength, which is
expected to cross zero when $L_3\approx M$.

    The Hamiltonian of the system can be derived in the standard way
\cite{Josephson-book,Kofman} and written in terms of the phases
$\phi_{i}$ ($i=1,2,3$) across the three Josephson junctions and the
conjugated momenta $p_{i}$ ($[\phi_i,p_j]=\imath\hbar \delta_{ij}$),
which are the corresponding node charges, multiplied by
$\tilde\Phi_0\equiv\Phi_0/2\pi=\hbar/2e$:
    \begin{eqnarray}
&& \hspace{-0.3cm}    H=
   \sum_{i=1}^3 \left[ \frac{p_i^2 }{2\tilde{C}_i \tilde\Phi_0^2} \,
   - E_{Ji}\cos\phi_i \right]
     \nonumber \\
&& +\sum_{i=1}^2 \frac{\tilde\Phi_0^2}{2L_i} \,
(\phi_i-\phi_{e,i})^2 - I_B \tilde\Phi_0 \phi_3
 +\frac{\tilde{C}_a}{C_1C_2}  \frac{p_1p_2}{\tilde\Phi_0^2}
    \nonumber \\
&& +\frac{\tilde{M}}{\tilde{L}_4\tilde{L}_5} \tilde\Phi_0^2
(\phi_1-\phi_3-\phi_{e,4})(\phi_2-\phi_3-\phi_{e,5})
   \nonumber \\
&& +\frac{\tilde\Phi_0^2}{2\tilde{L}_4} (\phi_1-\phi_3-\phi_{e,4})^2
+ \frac{\tilde\Phi_0^2}{2\tilde{L}_5} (\phi_2-\phi_3-\phi_{e,5})^2,
\qquad
    \label{Ham1}    \end{eqnarray}
where $\tilde{C}_1=C_1+C_2 C_a/(C_2+C_a)$ and $\tilde{C}_2=C_2+C_1
C_a/(C_1+C_a)$ are the effective qubit capacitances,
$\tilde{C}_3=C_3$ (introduced for notational convenience),
$\phi_{e,1}$ and $\phi_{e,2}$ are the external dimensionless qubit
fluxes, $\phi_{e,4}$ and $\phi_{e,5}$ are the external dimensionless
fluxes through the loops containing $L_4$ and $L_5$ (we will often
assume $\phi_{e,4}=\phi_{e,5}=0$),
 $\tilde{C}_a=(1/C_a+1/C_1+1/C_2)^{-1}$ is the effective
coupling capacitance, and renormalized coupling inductances are
    \begin{equation}
 \frac{\tilde{L}_4}{L_4}=\frac{\tilde{L}_5}{L_5}=\frac{\tilde{M}}{M}=
 1-\frac{M^2}{L_4 L_5}.
    \label{tilde-def}
    \end{equation}
 All terms in Eq.\ (\ref{Ham1}) have clear physical meaning.
\cite{magnetostatics-sign} Introducing the shifted variables $\delta
\phi_i=\phi_i-\phi_{i,\rm st}$, where the set $\{\phi_{i,\rm st}\}$
corresponds to the minimum of the potential energy, we rewrite the
Hamiltonian (\ref{Ham1}) as
    \begin{eqnarray}
&& \hspace{0.4cm}    H=H_1+H_2+H_3+H_{\rm int},
    \label{Ham-separ}\\
&&  \hspace{0.4cm}  H_i=\frac{p_i^2}{2\tilde{C}_i\tilde\Phi_0^2}  +
U_i(\delta \phi_i), \,\,\, i=1,2,3,
    \\
&& \hspace{-0.5cm} H_{\rm int}= \frac{\tilde{C}_a}{C_1C_2}
\frac{p_1p_2}{\tilde\Phi_0^2}
  + \tilde\Phi_0^2 \left(
  \frac{\tilde{M}}{\tilde{L}_4\tilde{L}_5}
    \delta\phi_1 \delta\phi_2 \right.
  \nonumber \\
  && \hspace{0.0cm}
\left. -\frac{1+\tilde{M}/\tilde{L}_5}{\tilde{L}_4}
    \delta\phi_1 \delta\phi_3
-\frac{1+\tilde{M}/\tilde{L}_4}{\tilde{L}_5}
    \delta\phi_2 \delta\phi_3 \right),
    \label{Ham-int}\end{eqnarray}
where the potentials $U_i$ have minima at $\delta\phi_i=0$, and the
corresponding plasma frequencies $\omega_{i,\rm
pl}=(\tilde{L}_i\tilde{C}_i)^{-1/2}$ are governed by the effective
inductances
  \begin{eqnarray}
&& \tilde{L}_1=[L_1^{-1}+\tilde{L}_4^{-1}+\tilde\Phi_0^{-2}
E_{J1}\cos
  \phi_{1,\rm{st}}]^{-1},
  \\
&& \tilde{L}_2=[L_2^{-1}+\tilde{L}_5^{-1}+\tilde\Phi_0^{-2}
E_{J2}\cos
  \phi_{2,\rm{st}}]^{-1},
  \\
&& \hspace{-0.0cm} \tilde{L}_3=\left[
\frac{1}{\tilde{L}_4}+\frac{1}{\tilde{L}_5}+
\frac{2\tilde{M}}{\tilde{L_4}\tilde{L}_5}+\frac{1}{\tilde\Phi_0}
\sqrt{I_{3,\rm cr}^2-I_{3,\rm st}^2} \right]^{-1}, \quad
  \end{eqnarray}
where in the last equation we expressed $E_{J3}\cos\phi_{3,\rm st}$
in terms of the corresponding junction current $I_{3,\rm st} \simeq
I_B$ and the critical current $I_{3,\rm cr}= E_{J3}/\tilde\Phi_0 $.

    The Hamiltonians $H_1$ and $H_2$ correspond to the separated
qubits; in absence of coupling two lowest eigenstates in each of
them correspond to the logic states $|0\rangle$ and $|1\rangle$.
Notice that because of an anharmonicity of the potentials, the qubit
frequencies $\omega_1$ and $\omega_2$ (defined via energy difference
between $|1\rangle$ and $|0\rangle$ for uncoupled qubits) are
slightly smaller than the plasma frequencies $\omega_{1,\rm pl}$ and
$\omega_{2,\rm pl}$. The coupler is characterized by the Hamiltonian
$H_3$; its similarly defined frequency $\omega_3$ is slightly
smaller than $\omega_{3,\rm pl}$. In the experiment \cite{Bialczak}
$\omega_3$ was almost an order of magnitude higher than
$\omega_{1,2}$, and the coupler had only virtual excitations. In our
analysis we also assume absence of real excitations in the coupler;
however, in general we do not assume $\omega_3\gg \omega_{1,2}$
(except specially mentioned), we only assume absence of resonance
between the coupler and the qubits.

\vspace{0.3cm}

    Our goal is to calculate the coupling between the two qubits
due to $H_{\rm int}$ (we assume a weak coupling). In general, the
coupling between two logic qubits can be characterized by 9 real
parameters. However, since for phase qubits the energies of the
states $|0\rangle$ and $|1\rangle$ are significantly different, we
can use the rotating wave approximation (RWA), i.e.\ neglect terms
creating two excitations or annihilating two excitations; then there
are only 3 coupling parameters in the rotating frame:
    \begin{eqnarray}
&& \hspace{-0.3cm}   H_{c}=\frac{\hbar \Omega_{XX}}{4} \left(
\sigma_{X}^{(1)}
    \sigma_{X}^{(2)}+\sigma_{Y}^{(1)}
    \sigma_{Y}^{(2)}\right)
  \nonumber \\
&& \hspace{0.3cm} +\frac{\hbar \Omega_{XY}}{4} \left(
\sigma_{X}^{(1)} \sigma_{Y}^{(2)} - \sigma_{Y}^{(1)}
\sigma_{X}^{(2)} \right) + \frac{\hbar \Omega_{ZZ}}{4}
\sigma_{Z}^{(1)} \sigma_{Z}^{(2)} , \qquad
    \label{H-general}\end{eqnarray}
where superscripts of the Pauli matrices indicate qubit numbering.
Moreover, from the symmetry of the interaction Hamiltonian
(\ref{Ham-int}) it follows that $\Omega_{XY}=0$ (because matrix
elements of $\delta \phi_i$ are real, and for $p_i$ they are
imaginary); therefore our goal is to calculate only two coupling
frequencies: $\Omega_{XX}$ and $\Omega_{ZZ}$. Notice that since the
$XX$ and $YY$ interactions are indistinguishable in the RWA [both
correspond to $\sigma^{(1)}_+\sigma^{(2)}_- +
\sigma^{(1)}_-\sigma^{(2)}_+$], we may rewrite the first term in
Eq.\ (\ref{H-general}) as $(\hbar\Omega_{XX}/2)\,\sigma_{X}^{(1)}
\sigma_{X}^{(2)}$.

    The considered system of Fig.\ \ref{Schematic} has a large
Hilbert space, which can in principle be reduced to a two-qubit
space in a variety of ways, giving in general different values of
$\Omega_{XX}$ and $\Omega_{ZZ}$. To avoid ambiguity, we $\it define$
the coupling frequencies in terms of the exact eigenstates of the
full physical system. In particular, we associate the two-qubit
logic states $|00\rangle$ and $|11\rangle$ with the corresponding
eigenstates of the full Hamiltomian (\ref{Ham-separ}) and denote
their energies as $E_{00}$ and $E_{11}$; for this association we
start with the product-state using the ground state for the coupler,
and then find the nearest eigenstate. Similarly, instead of trying
to define the uncoupled logic states $|01\rangle$ and $|10\rangle$
(that is ambiguous, though a natural definition can be based on the
dressed states discussed in Sec.\ IV), we deal with the eigenstates
resulting from their coupling, which are associated with two exact
eigenstates of the full system; their energies are denoted as $E_+$
and $E_-$ ($E_+\geq E_-$).

Then the coupling $\Omega_{ZZ}$ is defined as
   \be
\hbar \Omega_{ZZ}=E_{00}+E_{11}-(E_+ +E_-) ,
    \label{Omega-ZZ-def} \ee
which is obviously consistent with Eq.\ (\ref{H-general}) for logic
qubits (a similar definition has been used in Ref.\
\onlinecite{Yale-Nature-09}). The coupling $\Omega_{XX}$ can be
defined as the minimal splitting in the avoided level crossing
between $|01\rangle$ and $|10\rangle$, i.e. as
    \be
|\hbar\Omega_{XX}|=\min_{\omega_1-\omega_2}(E_+-E_-),
    \label{Omega-XX-def1}\ee
with the sign of $\Omega_{XX}$ easily obtained by comparing with
Eq.\ (\ref{H-general}). Actually, the definition
(\ref{Omega-XX-def1}) cannot be applied to an arbitrary qubit
detuning $\omega_1-\omega_2$ (in a symmetric case it works only for
degenerate qubits); such generalization of $\Omega_{XX}$ definition
can be done by comparing exact eigenstates with the standard avoided
level crossing behavior (discussed in more detail in Secs.\ IV and
V). Notice that our definitions of $\Omega_{XX}$ and $\Omega_{ZZ}$
do not need any assumption of a weak coupling (this is their main
advantage); however, a weak coupling will be assumed in derivation
of analytical results.

    \section{Simple semiclassical analysis}

    Let us first calculate $\Omega_{XX}$ and $\Omega_{ZZ}$ in a
simple, essentially electrical engineering way (we will see later
that the result is close to the quantum result). For simplicity in
this section we assume $(L_4,L_5)\gg
(\tilde{L}_{1},\tilde{L}_{2},\tilde{L}_{3},M)$, $C_a \ll (C_1,
C_2)$, so that the coupling is weak and the tilde signs in many
cases can be avoided. We also replace the middle junction with the
effective inductance $L_3=\tilde\Phi_0/\sqrt{I_{3,\rm cr}^2-I_{3,\rm
st}^2} \approx \tilde\Phi_0/\sqrt{I_{3,\rm cr}^2-I_B^2}$ (in this
approximation $\tilde L_3\approx L_3$).

    The coupling $\Omega_{XX}$ corresponds to the frequency
splitting between the symmetric and antisymmetric modes of the
two-qubit oscillations. So, let us assume degenerate qubits,
$\omega_1=\omega_2=\omega_{\rm qb}$, and find the splitting in the
classical linear system. Notice that at frequency $\omega_{\rm qb}$
the capacitance $C_3$ is equivalent to the inductance
$-1/(\omega_{\rm qb}^2C_3)$, and therefore the parallel connection
of $L_3$ and $C_3$ is equivalent to the inductance
   \be
L_{3}^{\rm eff}=\frac{L_3}{1-\omega_{\rm qb}^2 L_{3}C_3}= \frac{
L_{3}}{1-(\omega_{\rm qb}/\omega_3)^2};
    \label{L3^*}\ee
notice that here $\omega_3=\omega_{3,\rm pl}$ for the coupler since
we assume a linear system.

    Suppose $\omega\approx\omega_{\rm qb}$ is a classical eigenfrequency,
and the first qubit voltage is $V_1e^{i\omega t}$. Then using the
phasor representation, we find the current through $L_4$ as
$I_4=V_1/(i\omega L_4)$; it induces the voltage $V_{\rm cp}=i\omega
(L_{3}^{\rm eff}-M)I_4$ in the coupling inductances, which causes
the current $I_5=V_{\rm cp}/(i\omega L_5)=V_1 (L_{3}^{\rm
eff}-M)/(i\omega L_4L_5)$ flowing through $L_5$ into the second
qubit. Adding this current to the current $I_a=i\omega C_a V_1$
through $C_a$, we get the total current $I_2=I_5+I_a$, flowing into
the second qubit. The extra current $I_2$ is equivalent to changing
the qubit capacitance $C_2$ by $\Delta C_2=-I_2/(i\omega V_2)$,
where $V_2=\pm V_1\sqrt{C_1/C_2}$ is the second qubit voltage for
the symmetric and antisymmetric modes (the factor $\sqrt{C_1/C_2}$
comes from the condition of equal energies in the two qubits). The
effective change of the capacitance slightly changes the oscillation
frequency $(\tilde L_2 C_2)^{-1/2}$, so the eigenfrequency can be
found as $\omega = \omega_{\rm qb}(1-\Delta C_2/2C_2)$. Therefore,
the frequency splitting due to coupling is
$|\Omega_{XX}|=\omega_{\rm qb} |\Delta C_2|/C_2$, and substituting
$\Delta C_2$ we finally find
        \be
    \Omega_{XX}= \frac{M- L_3/ [1-(\omega_{\rm qb}/\omega_3)^2]}
    {L_4 L_5 \omega_{\rm qb} \sqrt{C_1C_2}}
 + \frac{C_a\, \omega_{\rm qb}}{\sqrt{C_1 C_2}} ,
    \label{Omega-XX-class}\ee
where the explicit expression (\ref{L3^*}) for $L_3^{\rm eff}$ has
been used, and the sign of $\Omega_{XX}$ is determined by noticing
that a positive $\Omega_{XX}$ should make the frequency (energy) of
the symmetric mode larger than for the antisymmetric mode [see Eq.\
(\ref{H-general})].

    The most important observation is that $\Omega_{XX}$ depends on
the bias current $I_B$, which changes $L_3$, and for a proper
biasing the coupling $\Omega_{XX}$ can be zeroed exactly. If the
correction due to the $C_a$-term is small and also $\omega_{\rm
qb}/\omega_3 \ll 1$ (as in the experiment \cite{Bialczak}), then
$\Omega_{XX}$ is zeroed when $L_3\approx M$.

    The coupling $\hbar\Omega_{ZZ}\sigma_Z^{(1)}\sigma_Z^{(2)}/4$
in Eq.\ (\ref{H-general}) originates from anharmonicity of the qubit
potentials and corresponding difference between the average
Josephson phases for states $|1\rangle$ and $|0\rangle$, which we
denote as $\Delta\phi_{10}^{(i)}$ for the $i$th qubit. This leads to
the extra dc current $\tilde\Phi_0\Delta\phi_{10}^{(1)}/L_4$ through
the inductance $L_4$, when the first qubit changes state from
$|0\rangle$ and $|1\rangle$, and similar dc current change
$\tilde\Phi_0\Delta\phi_{10}^{(2)}/L_5$ through $L_5$ for the second
qubit. As a result, the state $|11\rangle$ acquires an additional
magnetic interaction energy, which is the product of these two
currents multiplied \cite{magnetostatics-sign} by $M-L_3$. This
corresponds to
   \be
    \Omega_{ZZ}= \Delta\phi_{10}^{(1)}\Delta\phi_{10}^{(2)}
\frac{\tilde\Phi_0^2}{\hbar}
     \frac{M- L_3}{L_4 L_5} .
    \label{Omega-ZZ-class}\ee
Comparing this result with Eq.\ (\ref{Omega-XX-class}) for
$\Omega_{XX}$, we see absence of the contribution due to $C_a$
(which is small anyway) and a similar proportionality to $M-L_3$,
though without the correction $1-(\omega_{\rm qb}/\omega_3)^2$. This
means that by changing the bias current $I_B$ (which affects $L_3$),
the coupling $\Omega_{ZZ}$ can be zeroed, and this happens close to
the point where $\Omega_{XX}$ is zeroed. Except for the vicinity of
the crossing point, $|\Omega_{ZZ}/\Omega_{XX}| \ll 1$ because
$\Delta \phi_{10}^{(i)}$ is small (compared to the ground state
width) for a weak anharmonicity; therefore $\Omega_{XX}$ is the main
coupling in our system.

 \section{Quantum analysis (analytics)}

    For the quantum analysis let us rewrite the interaction
Hamiltonian (\ref{Ham-int}) as
    \begin{eqnarray}
&& \hspace{-0.3cm}    H_{\rm int} = K_{13} (a_1 +a_1^\dagger)(a_3
+a_3^\dagger) + K_{23} (a_2 +a_2^\dagger)(a_3+a_3^\dagger)
    \nonumber \\
&& + K_{12} (a_1+a_1^\dagger)(a_2+a_2^\dagger) +
 K_{12}^a (a_1 - a_1^\dagger)(a_2-a_2^\dagger), \qquad
    \label{H-int-aa+}    \end{eqnarray}
where $a_i+a_i^\dagger =\delta \phi_i\sqrt{2m_i\omega_i/\hbar}$,
$a_i-a_i^\dagger=\imath p_i\sqrt{2/\hbar m_i\omega_i}$, and
    \begin{eqnarray}
&& K_{13}=-\frac{1+\tilde{M}/\tilde{L}_5}{\tilde{L}_4} k_{13},
\,\,\, K_{23}=-\frac{1+\tilde{M}/\tilde{L}_4}{\tilde{L}_5} k_{23},
\qquad
     \label{K13,23}\\
&& K_{12}=\frac{\tilde{M}}{\tilde{L}_4\tilde{L}_5} k_{12}, \,\,\,
  K_{12}^a= - \frac{\tilde C_a}{C_1C_2} \frac{\hbar^2}{4k_{12}},
     \label{K12,12a} \\
&& k_{ij}= \frac{\hbar \tilde\Phi_0^2 }{2\sqrt{m_i\omega_i
m_j\omega_j}}, \,\,\, m_i = \tilde\Phi_0^2 \tilde C_i
     \label{kij,m}\end{eqnarray}
(we use the creation/annihilation operators $a_i^\dagger$ and $a_i$
only for brevity of notatons; in their normalization we
 use the frequency $\omega_i$ between two lowest eigenstates instead
 of the plasma frequency).

    In order to find $\Omega_{XX}$, we have to solve the Schr\"odinger
equation $H|\psi_\pm\rangle = E_\pm |\psi_\pm\rangle$ for the two
eingenstates $|\psi_\pm\rangle$ ($E_+\geq E_-$), corresponding to
the coupled logic states $|10\rangle$ and $|01\rangle$. The
wavefunction can be written in the product-state basis as $
|\psi_\pm\rangle =\alpha_\pm |100\rangle +\beta_\pm, |001\rangle +
... $, where in this notation we show the energy levels
$|n_1n_3n_2\rangle$  of the first qubit, the coupling oscillator (in
the middle), and the second qubit, and the terms not shown
explicitly should be relatively small in the weak coupling case.
Comparing the amplitudes and energies with the standard avoided
level crossing behavior, let us define two coupling frequencies,
$\Omega_{XX}^{\, +}$ and $\Omega_{XX}^{\, -}$, as
  \be
  \Omega_{XX}^{\,\pm}=\frac{E_\pm
-E_\mp}{\hbar}\, \frac{2\alpha_{\pm}/\beta_\pm
}{1+(\alpha_{\pm}/\beta_\pm)^2}.
    \label{Omega-XX-pm-def}\ee
We have to define two frequencies because in general
$\alpha_+/\beta_+ \neq -\beta_- /\alpha_-$, in contrast to the ideal
case of two logic qubits; this is the price to pay when the
two-qubit language is applied to a more complicated physical system.
However, the difference $|\Omega_{XX}^{\,+}-\Omega_{XX}^{\,-}|$ is
typically very small; moreover, for degenerate qubits in a symmetric
system $\Omega_{XX}^{\,+}=\Omega_{XX}^{\,-}$  (because then
$|\alpha_\pm/\beta_\pm|=1$), so that we need only a single frequency
$\Omega_{XX}$, which in this case coincides with the definition
(\ref{Omega-XX-def1}). We will neglect the difference between
$\Omega_{XX}^{\,\pm}$ and $\Omega_{XX}$ unless specially mentioned
(the difference is important in the case of strongly detuned
qubits).

  For the analysis it is convenient to express a solution of the
Schr\"odinger equation $H|\psi\rangle = E |\psi\rangle$ as $
|\psi\rangle =\alpha |\psi_{100}^{\rm dr}\rangle +\beta
|\psi_{001}^{\rm dr}\rangle$, where the dressed states
$|\psi_{100}^{\rm dr}\rangle$ and $|\psi_{001}^{\rm dr}\rangle$ are
defined in the following way. The state $|\psi_{100}^{\rm
dr}\rangle$ expanded in the product-state basis has the contribution
from the state $|100\rangle$ with amplitude 1 and zero contribution
from the state $|001\rangle$, i.e., $\langle \psi_{100}^{\rm dr}
|100\rangle =1$ and $\langle \psi_{100}^{\rm dr} |001\rangle =0$.
Also, $|\psi_{100}^{\rm dr}\rangle$ satisfies equation $\langle
n|H|\psi_{100}^{\rm dr}\rangle = E \langle n|\psi_{100}^{\rm
dr}\rangle$ for all basis elements
$|n\rangle\equiv|n_1n_3n_2\rangle$ except $|100\rangle$ and
$|001\rangle$. The dressed state $|\psi_{001}^{\rm dr}\rangle$ is
defined similarly, except now $\langle \psi_{001}^{\rm dr}
|100\rangle =0$ and $\langle \psi_{001}^{\rm dr} |001\rangle =1$.
Notice that a dressed state is not a solution of an eigenvalue
problem; for a given energy $E$ it is a solution of an inhomogeneous
systems of linear equations. Also notice that we do not need to
normalize the wavefunctions.

    Constructing the dressed states in this way, we have to satisfy
(self-consistently for $E$) only two remaining equations to solve
the Schr\"odinger equation:
    \begin{eqnarray}
    \alpha \langle 100 |H|\psi_{100}^{\rm dr}\rangle +
   \beta \langle 100 |H|\psi_{001}^{\rm dr}\rangle =
   E \alpha ,
   \label{two-lev-1}\\
    \alpha \langle 001 |H|\psi_{100}^{\rm dr}\rangle +
   \beta \langle 001 |H|\psi_{001}^{\rm dr}\rangle =
   E \beta .
    \label{two-lev-2}\end{eqnarray}
Using linear algebra it is easy to prove the reciprocity relation
$\langle 100 |H|\psi_{001}^{\rm dr}\rangle = \langle 001
|H|\psi_{100}^{\rm dr}\rangle^*$ (in our case the complex
conjugation is actually not needed since the matrix elements are
real), and therefore Eqs.\  (\ref{two-lev-1}) and (\ref{two-lev-2})
are similar to the standard equations for an avoided level crossing.
Hence, the matrix elements $E_{100}^{\rm dr}=\langle 100
|H|\psi_{100}^{\rm dr}\rangle$ and $E_{001}^{\rm dr}=\langle 001
|H|\psi_{001}^{\rm dr}\rangle$ play the role of renormalized
self-energies of the two-qubit logic states $|10\rangle$ and
$|01\rangle$, while the two-qubit coupling can be calculated as
    \be
     \Omega_{XX}= \frac{2}{\hbar}\langle 001 |H_{\rm int}|\psi_{100}^{\rm
     dr}\rangle = \frac{2}{\hbar}\langle 100 |H_{\rm int}|\psi_{001}^{\rm
     dr}\rangle ,
     \label{Omega-XX-quant-gen}\ee
so that the eigenenergies are given by the usual formula $E_\pm =
[E_{100}^{\rm dr}+E_{001}^{\rm dr} \pm \sqrt{(E_{100}^{\rm
dr}-E_{001}^{\rm dr})^2 +\hbar^2\Omega_{XX}^2}\,]/2$ [in Eq.\
(\ref{Omega-XX-quant-gen}) we wrote $H_{\rm int}$ instead of $H$
because there is obviously no contribution from the non-interacting
part]. Notice that the dressed states and therefore the matrix
elements depend on energy $E$, in contrast to the standard level
crossing. This leads to a slight difference of $\Omega_{XX}$ for the
eigenstates $E_+$ and $E_-$, and also makes calculations using Eq.\
(\ref{Omega-XX-quant-gen}) slightly different from the definition
(\ref{Omega-XX-pm-def}).

    To find $\Omega_{XX}$ analytically, let us assume that the three
oscillators are linear, $H_{i}=\hbar\omega_i (a_i^\dagger a_i
+1/2)$, and use the lowest-order perturbation theory.  In the first
order
   \begin{eqnarray}
&& |\psi_{100}^{\rm{dr}} \rangle = |100\rangle + K_{13} \left(
\frac{|010\rangle}{E -\epsilon_{010}} +\frac{\sqrt{2} \,
|210\rangle}{E-\epsilon_{210}} \right)
    \nonumber \\
&& \hspace{1cm} + K_{23}
 \frac{|111\rangle}{E-\epsilon_{111}}
 +(K_{12}+K_{12}^a)
 \frac{\sqrt{2}\, |201\rangle}{E-\epsilon_{201}} , \quad
    \label{dr100-first} \end{eqnarray}
where the energies $\epsilon$ of the basis states are only due to
non-interacting part $H_1+H_2+H_3$ of the Hamiltonian
(\ref{Ham-separ}). Then from Eq.\ (\ref{Omega-XX-quant-gen}) we
obtain
    \begin{equation}
    \Omega_{XX}=\frac{2}{\hbar} (K_{12}-K_{12}^a) + \frac{2}{\hbar}
    K_{13}K_{23} \left( \frac{1}{E-\epsilon_{010}}+\frac{1}{E-\epsilon_{111}}
    \right) .
    \label{Omega-XX-quant-lin1}\end{equation}
For degenerate qubits and weak coupling we can use approximation
$E\approx \epsilon_{100}=\epsilon_{001}$ in this equation, so that
$E-\epsilon_{010} \approx -\hbar (\omega_{3}-\omega_{\rm qb})$ and
$E-\epsilon_{111} \approx -\hbar(\omega_{3}+\omega_{\rm qb})$; then
using explicit expressions (\ref{K13,23})--(\ref{kij,m}) for the
matrix elements we finally obtain
        \be
    \Omega_{XX}= \frac{\displaystyle \tilde M- L_3^*/[
    1-(\omega_{\rm qb}/\omega_3)^2]}
    {\tilde L_4 \tilde L_5 \omega_{\rm qb} \sqrt{\tilde C_1 \tilde
    C_2}}
 + \frac{C_a^*\,
 \omega_{\rm qb}}{\sqrt{\tilde C_1 \tilde C_2}} ,
    \label{Omega-XX-quant}\ee
where $L_3^*=\tilde L_3 (1+\frac{M}{L_5})(1+\frac{M}{L_4})$, $C_a^*=
\tilde C_a (\tilde C_1 \tilde C_2/C_1 C_2)$, and we used
$\omega_3=[\tilde L_3 C_3]^{-1/2}$. Comparing this equation with the
classical result (\ref{Omega-XX-class}), we see that the results
coincide under assumptions used for the classical derivation.

   Validity of the perturbation theory requires assumptions
$|K_{13}/\hbar(\omega_3 \pm\omega_{\rm qb})|\ll 1$,
$|K_{23}/\hbar(\omega_3 \pm\omega_{\rm qb})|\ll 1$,
$|K_{12}/\hbar\omega_{\rm qb}|\ll 1$, and
$|K_{12}^a/\hbar\omega_{\rm qb}|\ll 1$, which basically mean that in
Eq.\ (\ref{Omega-XX-quant}) the contributions to $\Omega_{XX}$ due
to $\tilde M$ and $C_a^*$ should be much smaller than $\omega_{\rm
qb}$, and the contribution $L_3^*/\tilde L_4 \tilde L_5\omega_{\rm
qb}\sqrt{\tilde C_1\tilde C_2}$ should be much smaller than
$|\omega_{3}-\omega_{\rm qb}|^2/\omega_3$. Notice that we do not
need an assumption $\omega_{\rm qb}/\omega_{3}\ll 1$, we only need
absence of resonance between these frequencies; in fact, $\omega_3$
can be even smaller than $\omega_{\rm qb}$.

    To find analytics for $\Omega_{ZZ}$, it is necessary to
consider nonlinear oscillators, because this is what we expect from
the quasiclassical analysis and also because in the linear case for
degenerate qubits $\epsilon_{101}=\epsilon_{200}=\epsilon_{002}$,
and therefore there is an ambiguity in defining the logic state
$|11\rangle$.
 In order to calculate $\Omega_{ZZ}$ via
Eq.\ (\ref{Omega-ZZ-def}), we need to find the eigenenergies
$E_{00}$ and $E_{11}$, while calculation of $E_+ +E_- \approx
E_{100}^{\rm dr}+E_{001}^{\rm dr}$ has been already discussed above.
To find the ground state energy $E_{00}$ we introduce the dressed
state $|\psi_{000}^{\rm dr}\rangle$ as a state satisfying the
Schr\"odinger equation $\langle n|H|\psi\rangle = E \langle
n|\psi\rangle$ (for a given $E$) for all basis elements $|n\rangle$
except $|000\rangle$, and also satisfying condition $\langle
000|\psi_{000}^{\rm dr}\rangle =1$. In a similar way as above, we
construct $|\psi_{000}^{\rm dr}\rangle$ in the first order
perturbation theory and then find $E_{00}$ as $E_{000}^{\rm
dr}=\langle 000|H|\psi_{000}^{\rm dr}\rangle$, using approximation
$E\approx \epsilon_{000}$ in the construction of the dressed state.
To find the eigenenergy $E_{11}$, we introduce the dressed state
$|\psi_{101}^{\rm dr}\rangle$ in a similar way, then calculate it in
the first order, and then find $E_{11}$ as $E_{101}^{\rm dr}=\langle
101|H|\psi_{101}^{\rm dr}\rangle$ assuming $E\approx \epsilon_{101}$
for the dressed state.

    Even though this is a straightforward procedure, now there are
infinitely many terms in the first-order dressed states because of
the nonlinearity, and there are still many terms even if we keep
only lowest orders in nonlinearity. However, most of the
contributions to the energies cancel each other in the combination
$\hbar \Omega_{ZZ}=E_{000}^{\rm dr}+ E_{101}^{\rm dr}-E_{100}^{\rm
dr}-E_{001}^{\rm dr}$, and the largest non-canceling contributions
yield
    \begin{equation}
    \Omega_{ZZ} = \frac{b_1 b_2}{\hbar} \left[ K_{12} +
    \frac{2K_{13}K_{23}}{\epsilon_{101}-\epsilon_{111}}
     \right] ,
    \label{Omega-ZZ-quant1}\end{equation}
where $b_i$ is defined for $i$th oscillator as
  \begin{equation}
b_i =\frac{\langle 1|\delta \phi_i |1\rangle - \langle 0|\delta
\phi_i |0\rangle}{\sqrt{\hbar/2m_i\omega_i}} .
    \label{b-defin}\end{equation}
Notice that $\Omega_{ZZ}$ depends on the nonlinearity of qubits (via
$b_1$ and $b_2$), while nonlinearity of the coupling junction gives
only a small correction (see Appendix) to the second term in Eq.\
(\ref{Omega-ZZ-quant1}), which is neglected in the lowest order.
Also notice that in Eq.\ (\ref{Omega-ZZ-quant1}) we neglected terms
proportional to $K_{12}^2$, because they are on the same order as
the neglected terms $\sim (K_{13}K_{23})^2$. Using the definitions
(\ref{K13,23})--(\ref{kij,m}), we rewrite (\ref{Omega-ZZ-quant1}) in
the form
    \begin{equation}
   \Omega_{ZZ}= \frac{b_1 b_2}{2}\,
    \frac{\tilde{M}-
L_{3}^* }{\tilde{L}_4 \tilde{L}_5 \sqrt{\omega_1\omega_2}
\sqrt{\tilde C_1 \tilde C_2}} ,
    \label{Omega-ZZ-quant2}
    \end{equation}
which coincides with the classical result (\ref{Omega-ZZ-class})
under assumptions used for the classical result, since $\Delta
\phi_{10}^{(i)}=b_i\tilde\Phi_0^{-1}\sqrt{\hbar/2\tilde
C_i\omega_i}$.

    In deriving Eq.\ (\ref{Omega-ZZ-quant1}) we have used the lowest
order of the perturbation theory. However, there are higher-order
terms, which are significantly enhanced because the basis state
$|101\rangle$ is close to resonance with the states $|200\rangle$
and $|002\rangle$ even for degenerate qubits. Let us account for
this effect by analyzing the repulsion between these levels and
computing the corresponding shift of the eigenenergy $E_{11}$.
Following the above formalism for $\Omega_{XX}$, we find the level
splitting due to interaction between $|101\rangle$ and $|200\rangle$
to be $S_{|11\rangle ,|20\rangle}=2\langle 101|H_{\rm
int}|\psi_{200}^{\rm dr}\rangle$. Then writing
$|\psi_{200}^{\rm{dr}}\rangle$ in the same way as in Eq.\
(\ref{dr100-first}), we find $S_{|11\rangle ,|20\rangle}\approx
\sqrt{2}\hbar\Omega_{XX}$, which is essentially the same result as
for a qubit interacting with a resonator. \cite{Blais} Because of
the level repulsion, the eigenenergy $E_{11}$ has a shift by
$[\epsilon_{200}-\epsilon_{101} \pm
\sqrt{(\epsilon_{200}-\epsilon_{101})^2+2\hbar^2\Omega_{XX}^2}]/2$,
which in the dispersive case $\epsilon_{101}-\epsilon_{200}\gg
|\Omega_{XX}|$ becomes
$\hbar^2\Omega_{XX}^2/2(\epsilon_{101}-\epsilon_{200})$ (we assume
$\epsilon_{101}>\epsilon_{200}$, so the shift is up in energy). In
another notation
$\epsilon_{101}-\epsilon_{200}=\hbar(\omega_2-\omega_1+\delta\omega_1)$,
where by $\hbar\delta\omega_i$ we denote the correction for the
second excited level energy, $2\epsilon_1-\epsilon_0-\epsilon_2$,
for $i$th qubit. A similar shift up in energy for $E_{11}$ comes
from the interaction with the level $|002\rangle$. Adding these two
contributions, we modify Eq.\ (\ref{Omega-ZZ-quant2}) to become
   \begin{eqnarray}
&&    \Omega_{ZZ}=
 \frac{b_1 b_2}{2}\,
    \frac{\tilde{M}-
L_{3}^* }{\tilde{L}_4 \tilde{L}_5 \sqrt{\omega_1\omega_2}
\sqrt{\tilde C_1 \tilde C_2}}
    \nonumber \\
&& {\hspace{0.3cm}} +
 \frac{\Omega_{XX}^2}{2} \left(
\frac{1}{\omega_2-\omega_1+\delta\omega_1} +
\frac{1}{\omega_1-\omega_2+\delta\omega_2}  \right)
 . \qquad
    \label{Omega-ZZ-quant3}
    \end{eqnarray}
Notice that for $\Omega_{ZZ}$ we in general consider different qubit
frequencies $\omega_1$ and $\omega_2$, while in Eq.\
(\ref{Omega-XX-quant}) for $\Omega_{XX}$ we assumed nearly
degenerate qubits; however, unless qubit detuning significantly
affects $\Omega_{XX}$ (that will be discussed later), we can use
definition $\omega_{\rm qb}=(\omega_1+\omega_2)/2$ in Eq.\
(\ref{Omega-XX-quant}).

    Since $\Omega_{ZZ}$ has a major dependence on the qubit
nonlinearity, let us discuss it in more detail (see also Appendix).
  For $i$th oscillator potential with an additional cubic term,
it is convenient to characterize nonlinearity by the ratio
$N_i=U_{{\rm bar},i}/\hbar \omega_{i,\rm pl}$, where $U_{{\rm
bar},i}$ is the barrier height (assumed to be at $\delta \phi_j
>0$), so that $N_i$ is crudely the number of levels in the quantum
well ($N_{1,2}\sim 5$ in typical experiments with phase qubits
\cite{Cooper-04}). For a weak cubic nonlinearity ($N_i\gg 1$) one
can derive \cite{Martinis-bc} the following approximations:
    \begin{equation}
     b_i \approx 1/\sqrt{3N_i}, \,\,\,
\delta \omega_i \approx (5/36N_i)\, \omega_i .
    \label{b,dw}\end{equation}
Therefore, away from the point where $\Omega_{XX}\approx 0$, and
neglecting corrections due to non-zero $C_a$, $\omega_{\rm
qb}/\omega_3$, and $\omega_2-\omega_1$, the ratio of couplings is
    \begin{equation}
    \frac{\Omega_{ZZ}}{\Omega_{XX}} \approx \frac{1}{6\sqrt{N_1N_2}} +
    \frac{18 (N_1+N_2)}{5}\frac{\Omega_{XX}}{\omega_{\rm qb}} ,
    \end{equation}
which is quite small for typical experimental parameters. Notice
that for $N_i=5$ (which is typically used for qubits) the numerical
values $b_i=0.289$ and $\delta \omega_i /\omega_i=0.0378$ are
significantly different from what is expected from the large-$N$
analytics (\ref{b,dw}) (in the cubic approximation for the qubit
potential $b_i$ and $\delta \omega_i/\omega_i$ depend only on
$N_i$).

  \section{Residual coupling}

    Both $\Omega_{XX}$ and $\Omega_{ZZ}$ may cross zero when
$L_3^*$ is varied by adjusting the bias current $I_B$. However, they
are typically zeroed at different values of $L_3^*$, that prevents
turning the two-qubit coupling completely off. Since
$|\Omega_{ZZ}/\Omega_{XX}|\ll 1$ away from the zero-crossing points,
let us characterize the residual coupling by the $ZZ$-coupling value
$\Omega_{ZZ}^{\rm res}$ at the point where $\Omega_{XX}=0$.
    Using Eqs.\ (\ref{Omega-XX-quant}) and (\ref{Omega-ZZ-quant3}),
we find for degenerate qubits
   \begin{equation}
    \Omega_{ZZ}^{\rm res} \approx \frac{b_1 b_2}{2}
\left[    \frac{\tilde{M} \, (\omega_{qb} /\omega_3)^2}{\tilde{L}_4
\tilde{L}_5 \sqrt{\tilde C_1 \tilde C_2}\, \omega_{qb}}
  -\frac{C_a^* \omega_{\rm qb}}{\sqrt{\tilde C_1 \tilde C_2}} \right] ,
    \label{Omega-res}\end{equation}
where we assumed $\omega_{\rm qb}/\omega_3 \ll 1$, as in the
experiment.

It is natural to characterize the OFF/ON ratio for the adjustable
coupling by the ratio  $\Omega_{ZZ}^{\rm res} /\Omega_{XX}^{\rm
ON}$, where $\Omega_{XX}^{\rm ON}$ is the ``fully on''
$XX$-coupling. Let us assume the operating regime in which the
ON-coupling corresponds to zero bias current,\cite{exp-dif} and at
this point $\tilde M/L_3^* \simeq 2$ [see Eq.\
(\ref{Omega-XX-quant})]. Then $\Omega_{XX}^{\rm ON} \simeq \tilde
M/2\tilde L_4 \tilde L_5 \omega_{\rm qb}\sqrt{\tilde C_1\tilde
C_2}$, and we find an estimate
   \begin{equation}
    \Omega_{ZZ}^{\rm res} /\Omega_{XX}^{\rm ON} \simeq b_1
  b_2
\left[
   (\omega_{qb}/\omega_3 )^2
  - C_a^* \omega_{\rm qb}^2 \tilde L_4 \tilde L_5 /\tilde M \right]
  .
    \label{off-on}\end{equation}
 In particular, for
$C_a=0$, $N_1=N_2=5$, and $\omega_{\rm qb}/\omega_3 \approx 1/5$
(typical experimental parameters), this gives OFF/ON$\,\,\simeq
3\times 10^{-3}$. The capacitance $C_a\simeq M/L_4 L_5\omega_3^2$
needed to zero the OFF/ON ratio is then around 0.6 fF for typical
experimental parameters $L_{4,5}\simeq 3$ nH, $M\simeq 200$ pH,
$\omega_3/2\pi \simeq 30$ GHz.

    Actually, since our analytics is only the leading-order
calculation, while in $\Omega_{ZZ}^{\rm res}$ we have an almost
exact cancelation of contributions due to $\tilde M$ and $L_3^*$, we
cannot expect that Eqs.\ (\ref{Omega-res}) and (\ref{off-on}) are
accurate even in the leading order. Nevertheless, we can trust the
result that the OFF/ON ratio is typically quite small, because both
$b_1 b_2\approx (3\sqrt{N_1 N_2})^{-1}$ and the terms in the
brackets in Eq.\ (\ref{off-on}) are small. Moreover, the OFF/ON
ratio can be made exactly zero by choosing proper values for
$\omega_{\rm qb}/\omega_3$ and $C_a$. Even if our analytics missed a
small term in Eq.\ (\ref{off-on}), the OFF/ON ratio can be zeroed
either by increasing $C_a$ or decreasing $\omega_3$, since this
moves $\Omega_{ZZ}^{\rm res}$ in the opposite directions.

    Since $\Omega_{ZZ}^{\rm res}/\Omega_{XX}^{\rm ON}$ is very small or
even zero, we have to carefully consider other effects, which do not
vanish when both $\Omega_{XX}$ and $\Omega_{ZZ}$ discussed above are
zero. One of such effects becomes clear when we consider the case of
a strong qubit detuning, $|\omega_1-\omega_2|\gg |\Omega_{XX}|$. In
this case one would expect that the eigenstate close to the state
$|100\rangle$ should have a negligible contribution of the state
$|001\rangle$ and vise versa (so that the logic states $|10\rangle$
and $|01\rangle$ are decoupled); however, actually these
contributions cannot be decreased to zero. As seen from Eq.\
(\ref{Omega-XX-quant-lin1}), $\Omega_{XX}$ depends on energy $E$,
and therefore it is slightly different  for the two eigenstates with
energies $E_+$ and $E_-$. The difference $\Delta
\Omega_{XX}=\Omega_{XX}^{\, +}-\Omega_{XX}^{\, -}$ is approximately
$ -4K_{13}K_{23}(E_+ -E_-)/\hbar^3 \omega_3^2$, which can be
rewritten as
          \begin{equation}
    \Delta \Omega_{XX} \approx \frac{-L_3^* \, |\omega_1-\omega_2|}{
\tilde L_4 \tilde L_5 \omega_{\rm qb} \sqrt{\tilde C_1 \tilde
    C_2}\, \omega_3}
     \simeq -2\Omega_{XX}^{\rm ON}
    \frac{|\omega_1-\omega_2|}{\omega_3} ,
    \label{Delta-Omega-XX}\end{equation}
where we assumed $\omega_3 \gg \omega_{1,2}$ and used $L_3^* \approx
\tilde M$ for nearly OFF coupling (while $\tilde M/L_3^* \simeq 2$
in the ON regime at $I_B=0$). This means that if we zero the
amplitude of the state $|100\rangle$ in one of the eigenstates,
there will still be a non-zero amplitude of the state $|001\rangle$
in the other eigenstate (in contrast to an ideal two-qubit
situation). Choosing the smallest $\Omega_{XX}$ coupling as $\pm
\Delta\Omega_{XX}/2$, we obtain contributions $-\Omega_{XX}^{\rm
ON}/2\omega_3$ of the wrong states in both eigenstates. So, the
error occupation is $\sim (\Omega_{XX}^{\rm ON}/2\omega_3)^2$, which
for typical parameters is around $10^{-6}$.

    Another effect, which is related to inaccuracy of the RWA
approximation, can be characterized by the contribution of the state
$|101\rangle$ in the ground state. Using the second-order
perturbation theory for the dressed state $|\psi_{000}^{\rm
dr}\rangle$ we find
    \begin{equation}
\langle 101 |\psi_{000}^{\rm dr}\rangle = \frac{
K_{12}+K_{12}^a+\frac{K_{13}K_{23}}{\epsilon_{000}-\epsilon_{110}}
+\frac{K_{13}K_{23}}{\epsilon_{000}-\epsilon_{011}}}{\epsilon_{000}-\epsilon_{101}},
    \label{101-000} \end{equation}
which is approximately $-\Omega_{XX}/4\omega_{qb}$ when
$\Omega_{XX}$ is not close to zero, exactly as expected for the
non-RWA contribution $(\hbar\Omega_{XX}/2)
\sigma_+^{(1)}\sigma_+^{(2)}$ from the term $(\hbar\Omega_{XX}/2)
\sigma_X^{(1)}\sigma_X^{(2)}$ in the two-qubit Hamiltonian. However,
we are mostly interested in the case $\Omega_{XX}=0$; then Eq.\
(\ref{101-000}) becomes approximately $-\Omega_{XX}^{\rm
ON}/2\omega_3$, and the corresponding error occupation is $\sim
(\Omega_{XX}^{\rm ON}/2\omega_3)^2 \simeq 10^{-6}$, same as for the
strong-detuning effect.

   \section{Numerical results}

    To calculate $\Omega_{XX}$ and $\Omega_{ZZ}$ numerically, we
first find the phases $\{\phi_{1,\rm st}, \phi_{2,\rm
st},\phi_{3,\rm st} \}$, which correspond to the minimum potential
energy in Eq.\ (\ref{Ham1}), and then find the eigenfunctions and
eigenenergies of the Hamiltonian (\ref{Ham-separ})--(\ref{Ham-int})
using the product-state basis of energy levels in the three
anharmonic oscillators (in this calculation we use the cubic
approximation for the oscillator potentials). After that the
coupling $\Omega_{ZZ}$ is calculated from the eigenenergies using
Eq.\ (\ref{Omega-ZZ-def}), while for $\Omega_{XX}$ we calculate two
values, $\Omega_{XX}^{\,+}$ and $\Omega_{XX}^{\,-}$, using the
definition (\ref{Omega-XX-pm-def}). However, for the case of
degenerate qubits, which we mostly consider below, there is no
difference between $\Omega_{XX}^{\,+}$ and $\Omega_{XX}^{\,-}$.

\begin{figure}[tb]
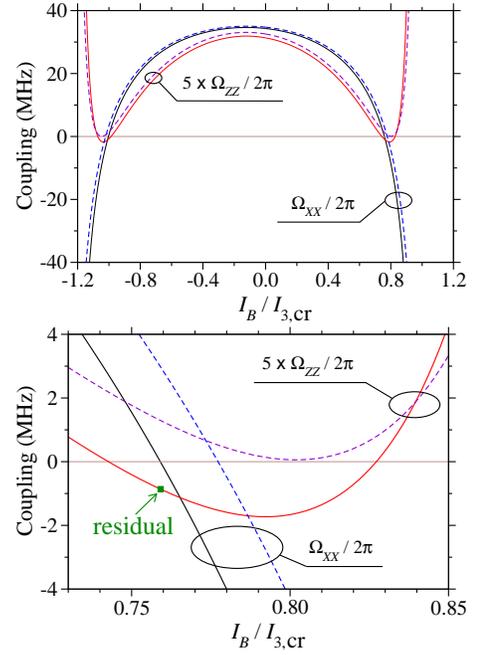

  \centering
\hspace{-0.0cm}\includegraphics[width=6.0cm]{tn-fg2a.eps}
\hspace{-0.0cm}\includegraphics[width=6.0cm]{tn-fg2b.eps}
  \caption{The coupling $\Omega_{XX}/2\pi$ and $\Omega_{ZZ}/2\pi$
(multiplied by 5 for clarity) as functions of the bias current
$I_B$. Solid lines show numerical results, while dashed lines show
analytics using Eqs.\ (\ref{Omega-XX-quant}) and
(\ref{Omega-ZZ-quant3}). The circuit parameters are: $I_{1,\rm
cr}=I_{2,\rm cr}=1.5 \, \mu A$, $I_{3,\rm cr}=3 \, \mu$A, $C_1=C_2=
1\,$pF, $C_3=0.1\,$pF, $C_a=0$, $L_1=L_2=0.7\,$nH, $L_4=L_5=3\,$nH,
$M=0.2\,$nH; $\phi_{e,4}=\phi_{e,5}=0$, $N_1=N_2=5$ ($\omega_1/2\pi
=\omega_2/2\pi=6.59\,$GHz). The panel (b) is a blow-up of the panel
(a) near the crossing points at positive $I_B$. We see that
$\Omega_{XX}=0$ at $I_B/I_{3,\rm cr}=0.759$, and the residual
coupling (square symbol) is $\Omega_{ZZ}^{\rm res}/2\pi=-172\,$kHz.
  }
  \label{fig2}
\end{figure}

    Figure \ref{fig2}(a) shows $\Omega_{XX}$ and $\Omega_{ZZ}$ as
functions of the bias current $I_B$ for the system with the
following parameters:\cite{exp-dif} $I_{1,\rm cr}=I_{2,\rm cr}=1.5
\, \mu A$, $I_{3,\rm cr}=3 \, \mu$A, $C_1=C_2= 1\,$pF,
$C_3=0.1\,$pF, $C_a=0$, $L_1=L_2=0.7\,$nH, $L_4=L_5=3\,$nH,
$M=0.2\,$nH. We assume $\phi_{e,4}=\phi_{e,5}=0$ for the coupler
loops, while the qubit external fluxes $\phi_{e,1}=\phi_{e,2}$ are
chosen so that for the qubits $N_1=N_2=5$; this corresponds to the
qubit frequencies $\omega_1/2\pi =\omega_2/2\pi=6.59\,$GHz, which
are kept constant with changing $I_B$ by the compensating change of
external fluxes $\phi_{e,1}$ and $\phi_{e,2}$. The results for
$\Omega_{ZZ}$ are multiplied by 5 for clarity (to become visually
comparable to $\Omega_{XX}$). The solid lines in Fig.\ \ref{fig2}(a)
show numerical results, while dashed lines show the analytics using
Eqs.\ (\ref{Omega-XX-quant}) and (\ref{Omega-ZZ-quant3}). One can
see that overall the analytics gives a pretty good approximation.
Smaller numerical value for $\Omega_{XX}$ than in analytics can be
partially explained by the corrections shown in Eqs.\
(\ref{Omega-XX-nonlin1}) and (\ref{Omega-XX-nonlin2}) in Appendix.
We have checked numerically that the beating frequency of the
classical small-amplitude oscillations is close to the analytical
quantum result for $\Omega_{XX}$ shown in Fig.\ \ref{fig2}(a), with
a typical difference on the order of 1 MHz.

    The lines in Fig.\  \ref{fig2}(a) are not symmetric about
$I_B=0$ because of the current through the coupling junction coming
from the qubits, which adds to $I_B$ (the curves are symmetric about
$I_B=-0.122\, I_{3,\rm cr}$; this asymmetry could be removed if the
qubits are biased with opposite fluxes, $\phi_{e,2}\approx
-\phi_{e,1}$, so that the currents from the qubits compensate each
other). At zero bias (ON-coupling) the coupling $\Omega_{XX}/2\pi$
is 34.3 MHz, which (analytically) comes from 85.8 MHz coupling due
to the mutual inductance $-M$ and compensating $-51.1$ MHz from the
inductance $\tilde L_3$ of the coupling junction (analytical total
slightly differs from the numerical result).

 At both positive and negative
$I_B$ the coupling $\Omega_{XX}$ crosses zero because of increase of
$\tilde L_3$, while $\Omega_{ZZ}$ barely crosses zero because of the
similar increase of $\tilde L_3$ and always positive contribution
from the level repulsion effect [see Eq.\ (\ref{Omega-ZZ-quant3})].
Figure \ref{fig2}(b) is a blow-up of \ref{fig2}(a) near the crossing
points at positive $I_B$. One can see that numerically calculated
residual coupling (at $\Omega_{XX}=0$) is $\Omega_{ZZ}^{\rm
res}/2\pi=-172\,$kHz, so that the OFF/ON ratio is $5\times 10^{-3}$.
While this value of $\Omega_{ZZ}^{\rm res}$ is on the same order as
expected from the analytics (see dashed lines), it has the opposite
(negative) sign. This apparently happens because corrections to
analytics in this case have a stronger effect than the effect of the
term $(\omega_{\rm qb}/\omega_3)^2$ in Eq.\ (\ref{Omega-XX-quant});
at the point where $\Omega_{XX}=0$ we have
$\omega_3/2\pi=38.6\,$GHz, so $(\omega_{\rm qb}/\omega_3)^2=0.029$
is really quite small ($\omega_3/2\pi= 49.6\,$GHz at $I_B=0$).
Notice that a negative value of $\Omega^{\rm res}_{ZZ}$ makes
impossible to zero $\Omega^{\rm res}_{ZZ}$ by adding the capacitive
coupling via $C_a$.

\begin{figure}[tb]
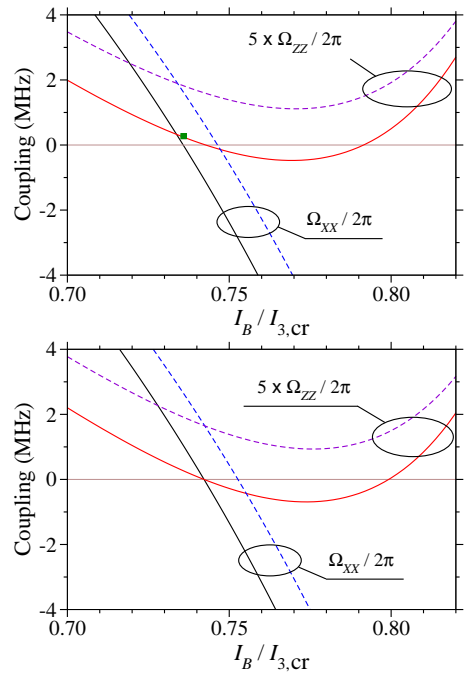

  \centering
\hspace{-0.0cm}\includegraphics[width=6.0cm]{tn-fg3a.eps}
\hspace{-0.0cm}\includegraphics[width=6.0cm]{tn-fg3b.eps}
  \caption{(a) Same as in Fig.\ \ref{fig2}(b), but for a circuit
with $C_3=0.3\,$pF. This makes positive $\Omega_{ZZ}^{\rm
res}/2\pi=49\,$kHz (square symbol). (b) Same as in (a), but with
added coupling capacitance $C_a=0.155\,$pF, that produces
$\Omega_{ZZ}^{\rm res}=0$, i.e.\ the couplings $\Omega_{XX}$ and
$\Omega_{ZZ}$ are zeroed simultaneously.}
  \label{fig3}
\end{figure}

    In order to make $\Omega^{\rm res}_{ZZ}$ positive, we can
decrease $\omega_3$ by increasing the coupling junction capacitance
$C_3$. Figure \ref{fig3}(a) shows results for the same circuit with
$C_3=0.3\,$pF, which is 3 times larger than in Fig.\ \ref{fig2}. We
show only vicinity of the crossing points, while the overall shape
of the curves is quite close to what is shown in Fig.\
\ref{fig2}(a); in particular, $\Omega_{XX}/2\pi=32.5\,$MHz at
$I_B=0$. As we see from Fig.\ \ref{fig3}(a), now $\Omega_{ZZ}^{\rm
res}$ becomes positive, $\Omega_{ZZ}^{\rm res}/2\pi=49\,$kHz (at
this crossing point $\omega_3/2\pi=22.9\,$GHz, and it is 28.7 GHz at
$I_B=0$). The corresponding OFF/ON ratio is now $1.5\times 10^{-3}$.

    Obviously, $\Omega_{ZZ}^{\rm res}=0$ for some intermediate value
of $C_3$. We have calculated that it happens for $C_3=0.253\,$pF
(keeping other parameters unchanged). If $C_3$ is larger than this
value, so that $\Omega_{ZZ}^{\rm res}$ is positive, we can zero the
residual coupling by adding small coupling capacitance $C_a$. Figure
\ref{fig3}(b) shows the results for $C_3=0.3\,$pF and
$C_a=0.155\,$fF (other parameters unchanged), in which case
$\Omega_{ZZ}^{\rm res}=0$.

\begin{figure}[tb]
  \centering
\hspace{-0.0cm}\includegraphics[width=6.0cm]{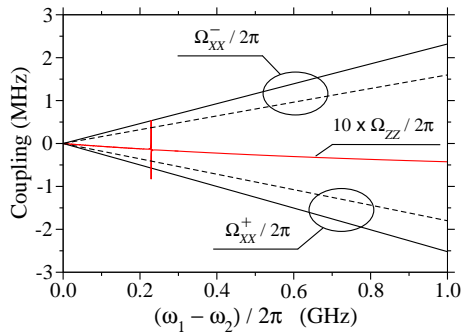}
  \caption{Solid lines: numerical results for $\Omega_{XX}^{\, +}$,
$\Omega_{XX}^{\, -}$, and $\Omega_{ZZ}$ (multiplied by 10) as
functions of the qubit detuning $\omega_1-\omega_2$ for the
parameters of Fig.\ \ref{fig3}(b) at the point where
$\Omega_{XX}=\Omega_{ZZ}=0$. Dashed lines show the analytics for
$\Omega_{XX}^{\, \pm} \approx \pm \Delta \Omega_{XX}/2$ from Eq.\
(\ref{Delta-Omega-XX}). Error state occupation due to nonzero
$\Omega_{XX}^{\,\pm}$ is $1.5\times 10^{-6}$.}
  \label{fig4}
\end{figure}

    At the point where $\Omega_{XX}=\Omega_{ZZ}=0$, we have to pay a
special attention to other effects which couple the two qubits. In
particular, we should consider what happens when the qubits are
detuned in frequency. Figure \ref{fig4} shows $\Omega_{XX}^{\,+}$,
$\Omega_{XX}^{\,-}$, and $\Omega_{ZZ}$ (multiplied by 10) as
functions of the detuning $\omega_1-\omega_2$ for the circuit with
$C_3=0.3\,$pF and $C_a=0.155\,$fF (other parameters as above) at the
bias current $I_B=0.742\,I_{3,\rm cr}$; these parameters correspond
to the point $\Omega_{XX}=\Omega_{ZZ}=0$ in Fig.\ \ref{fig3}(b). For
the detuning we change the external fluxes $\phi_{e,1}$ and
$\phi_{e,2}$ (and correspondingly change $N_1$ and $N_2$), while
keeping the frequency $\omega_{\rm qb}=(\omega_1+\omega_2)/2$
unchanged (and we still assume $\phi_{e,4}=\phi_{e,5}=0$). As seen
in Fig.\ \ref{fig4}, the couplings $\Omega_{XX}^{\,+}$ and
$\Omega_{XX}^{\,-}$ coincide when $\omega_1=\omega_2$; however,
their difference grows with the qubit detuning. The dashed lines
show the analytics
 for $\pm \Delta \Omega_{XX}/2$ using the first expression in Eq.\
(\ref{Delta-Omega-XX}). There is a significant difference between
the analytical and numerical results because the ratio $\omega_{\rm
qb}/\omega_3=6.59\,$GHz$/23.0\,$GHz is not very small; the next
order correction to the analytics by the factor  $[1+3(\omega_{\rm
qb}/\omega_3)^2]$,  which comes from Eq.\
(\ref{Omega-XX-quant-lin1}), accounts for most of the difference.
 Even though $|\Omega_{XX}^{\,\pm}|$ grows with the
detuning, the corresponding error state occupation
$|\alpha/\beta|^{\pm 2}$ is constant and is only $1.5\times
10^{-6}$.
  The detuning also changes $\Omega_{ZZ}$; however, the effect is
minor, and $\Omega_{ZZ}/2\pi =-42.6\,$kHz at the detuning of
$1\,$GHz. An almost vertical feature on the $\Omega_{ZZ}$ line at
the detuning of 0.23 GHz is due to the level crossing between states
$|101\rangle$ and $|002\rangle$. It is relatively small because we
have chosen the operating point with $\Omega_{XX}=0$ in absence of
detuning; otherwise at the level crossing point the coupling
$\Omega_{ZZ}$ changes by approximately $\pm \Omega_{XX}/ \sqrt{2}$
[see discussion above Eq.\ (\ref{Omega-ZZ-quant3})].

    For the numerical results in this section we have used the cubic
approximation for the qubit and coupling oscillator potentials in
calculation of the matrix elements of the Hamiltonian. We have also
done calculations using the exact potential and checked that the
values of $\Omega_{XX}$ and $\Omega_{ZZ}$ change only slightly,
though the residual coupling changes more significantly as expected
for an almost exact cancelation of contributions with opposite
signs.

  \section{Conclusion}

The main goal of this paper has been calculation of the two-qubit
coupling frequencies $\Omega_{XX}$ and $\Omega_{ZZ}$ for the circuit
of Fig.\ \ref{Schematic}, and analysis of their dependence on the
bias current $I_B$, which can be used to turn the coupling on and
off.\cite{Bialczak} We have shown that a simple ``electrical
engineering'' analysis for $\Omega_{XX}$ as the beating frequency of
two classical oscillators gives a result,  Eq.\
(\ref{Omega-XX-class}), which is close to the analytical
(lowest-order) quantum result, Eq.\ (\ref{Omega-XX-quant}) (the
formulas coincide, except minor renormalization of parameters). In
turn, the quantum analytics for $\Omega_{XX}$ is close to the
results of the quantum numerical analysis (Fig.\ \ref{fig2}); a
minor difference is due to higher orders in perturbation. The
``electrical engineering'' analysis for $\Omega_{ZZ}$, Eq.\
(\ref{Omega-ZZ-class}), is not fully classical; it needs the
language of quantum energy levels and shows that $\Omega_{ZZ}$
originates due to anharmonicity of the qubit oscillators, which
shifts the average flux. However, this analysis corresponds to only
one term in the quantum analytics, Eq.\ (\ref{Omega-ZZ-quant3}),
while the other significant contribution is due to the level
repulsion between the states with the single-excitation in each
qubit and with the double-excitation in one of them. The results of
the numerical quantum analysis for $\Omega_{ZZ}$ are similar to the
quantum analytics (Fig.\ \ref{fig2}).

As expected, our analysis shows that $\Omega_{XX}$ is the main
two-qubit coupling in the considered circuit, and $\Omega_{ZZ}$ is
typically much smaller. Nevertheless, in the analyzed numerical
example using realistic experimental parameters, the ratio
$\Omega_{XX}/\Omega_{ZZ}$ is only around 5 for the coupling turned
on (small $I_B$), which means that corrections for non-zero
$\Omega_{ZZ}$ in experimental algorithms are necessary.

    The most practically important case is when the coupling is
almost off. The fact that $\Omega_{XX}$ can be zeroed exactly is
rather trivial: a real number changing sign should necessarily cross
zero, and $\Omega_{XX}$ obviously changes sign when the effect of
the coupling junction inductance $L_3$ overcompensated the effect of
the magnetic coupling $-M$. The coupling $\Omega_{ZZ}$ does not
necessarily change sign because of always positive contribution from
the level repulsion, but $\Omega_{ZZ}$ should be small when
$\Omega_{XX}=0$, as follows from the analytical result
(\ref{Omega-ZZ-quant3}). This fact is quite beneficial for the
ability to turn the coupling almost off, and we have defined the
residual coupling $\Omega_{ZZ}^{\rm res}$ as the value of
$\Omega_{ZZ}$ when $\Omega_{XX}=0$. A natural measure of the
coupling OFF/ON ratio is $\Omega_{ZZ}^{\rm res}/\Omega_{XX}^{\rm
ON}$, where $\Omega_{XX}^{\rm ON}$ corresponds to the case of small
(zero) $I_B$. Notice that this ratio depends on the definition
(\ref{H-general}) since we compare different couplings (for example,
if the $ZZ$-term was defined in (\ref{H-general}) without the factor
4, then the OFF/ON ratio would decrease 4 times). As we found from
the analytical and numerical calculations, the OFF/ON ratio is few
times $10^{-3}$ for typical experimental parameters. Most
importantly, the OFF/ON ratio can be zeroed exactly by properly
choosing capacitances $C_3$ and/or $C_a$.

    Even when the above defined OFF/ON ratio is exactly zero, this
does not mean that the qubits can be made completely decoupled. The
reason is that the qubits are still physically coupled to the
coupling circuit, and the discussion of such effects should go
beyond the language of the coupling of logical qubits, which is
characterized by only $\Omega_{XX}$ and $\Omega_{ZZ}$. In
particular, detuning of the formally decoupled qubits leads to an
erroneous state occupation of around $(\Omega_{XX}^{\rm
ON}/2\omega_3)^2$, which is on the order of $10^{-6}$ for typical
experimental parameters. Non-RWA corrections bring errors of the
same order.

    In this paper we sometimes assumed a high frequency of the coupling
oscillator, $\omega_{3}\gg\omega_{\rm qb}$, as in the experiment.
\cite{Bialczak} However, the analytical results
(\ref{Omega-XX-quant}) and (\ref{Omega-ZZ-quant3}), which do not
rely on this assumption, show that it is not really needed for the
operation of the scheme. Moreover, the adjustable coupling can be
even realized for $\omega_{3}< \omega_{\rm qb}$; however in that
case we have to use positive mutual inductance, and we should not
expect typically small $\Omega_{ZZ}$ when $\Omega_{XX}=0$.

    The calculation of the coupling $\Omega_{XX}$ and $\Omega_{ZZ}$ in
this paper is based on the analysis of the eigenfunctions and
eigenenergies of the whole system, thus avoiding ambiguity of
reducing the whole system to two logical qubits. However, we have
also done the analytical calculations by using such reduction.
Assuming $\omega_{\rm qb}/\omega_3 \ll 1$, we have eliminated the
coupling junction degree of freedom by applying the Schrieffer-Wolf
transformation and then projecting the resulting Hamiltonian on the
coupling junction ground state; after that the Hamiltonian has been
truncated to the two-qubit subspace. The obtained results for
$\Omega_{XX}$ and $\Omega_{ZZ}$ basically coincide with Eqs.\
(\ref{Omega-XX-quant}) and (\ref{Omega-ZZ-quant2}) under the
assumptions used.

    The analyzed tuneable coupler (without $C_a$) has been realized
by Bialczak et al.,\cite{Bialczak} and the dependence of the
coupling frequency $\Omega_{XX}$ on the bias current $I_B$ has been
measured experimentally (the coupling $\Omega_{ZZ}$ has not been
measured). Due to a relatively small critical current $I_{3,\rm cr}$
of the coupling junction in the experiment, the coupling
$\Omega_{XX}$ is crossing zero at small $I_B$ (see Fig.\ 4d of Ref.\
\onlinecite{Bialczak}). Therefore, in contrast to the case shown in
our Figs.\ 2 and 3, the experimental coupler is nearly OFF at
$I_B=0$.  Notice that Fig.\ 4d of Ref.\ \onlinecite{Bialczak} shows
$-\Omega_{XX}$ (in our notation), so it increases with $|I_B|$, and
the dependence on $I_B$ is symmetrized by a horizontal shift. Using
the experimental parameters, we have checked that the theoretical
result for $\Omega_{XX}$ is quite close to the experimental result
(Fig.\ 4d of Ref.\ \onlinecite{Bialczak} shows a fitting by the
simple theory, which is close to the full theory result).

    For realization of multi-qubit algorithms it is very important
that the residual coupling $\Omega_{ZZ}^{\rm res}$ can be zeroed by
proper design of $C_3$ and $C_a$. If this is not done, the OFF/ON
ratio is small (few times $10^{-3}$), but may still be significant
for complicated algorithms. There is a modification of the scheme of
Fig.\ \ref{Schematic}, which may further reduce the OFF/ON ratio
without using $C_a$ and without precise choice of $C_3$. The idea is
to add blocking capacitors between the qubits and inductors
$L_{4,5}$. Then there will be no dc current from the qubits going
into the coupling circuit, and this will eliminate the classical
interaction effect leading to $\Omega_{ZZ}$ in Eq.\
(\ref{Omega-ZZ-class}). Correspondingly, this should eliminate the
first term in the quantum result (\ref{Omega-ZZ-quant3}) for
$\Omega_{ZZ}$, so that $\Omega_{ZZ}^{\rm res}$ should be very small
by itself. In the quantum language, this happens because in the
modified scheme the capacitive interaction between five oscillators
is of the momentum-momentum type (besides one phase-phase
interaction due to $M$), and the average momentum for any eigenstate
of an oscillator is exactly zero. From experimental point of view,
the scheme with blocking capacitors is more convenient because it
eliminates the need to adjust external fluxes in the qubits when the
bias current $I_B$ is changed. We have performed preliminary quantum
calculations for $\Omega_{XX}$, which confirm that $\Omega_{XX}$
crosses zero when the effective inductance $L_3^{\rm eff}$
compensates the magnetic coupling $-\tilde M$, similar to the case
without blocking capacitors. However, mathematically this involves
compensation of three dozen quantum terms of the same order, so we
may expect that the scheme with blocking capacitors is less robust
against decoherence than the scheme of Fig.\ \ref{Schematic}. Such a
comparative analysis of the schemes with and without blocking
capacitors is a subject of further study.

This work was supported by NSA and IARPA under ARO grant
W911NF-08-1-0336.

\appendix*
\section{Corrections due to nonlinearity}

In this appendix we discuss an oscillator with a weak cubic
nonlinearity, and show next-order corrections for $\Omega_{XX}$ and
$\Omega_{ZZ}$ due to nonlinearity.

    Let us consider an oscillator with a cubic nonlinearity,
    $H=p^2/2m+(m\omega_{\rm pl}^2/2) (\delta\phi)^2-\lambda
(\delta\phi)^3$, where $\omega_{\rm pl}$ is the plasma frequency and
$\lambda >0$, so that there is a finite barrier height $U_{\rm
bar}=m^3\omega_{\rm pl}^6/54\lambda^2$ at positive $\delta\phi$. It
is convenient to characterize nonlinearity by the ratio $N=U_{\rm
bar}/\hbar \omega_{\rm pl}$, so that $N$ is crudely the number of
levels in the quantum well.

    Nonlinearity changes the eigenstates $|k\rangle$, eigenenergies
$\epsilon_k = \langle k|H|k\rangle$, and the normalized matrix
elements of the coordinate and momentum operators,
    \begin{equation}
 c_{kl}=c_{lk}=\frac{\langle k |\delta\phi|l\rangle }{
\sqrt{\hbar/2m\omega_{\rm pl}}} , \,\,\,
d_{kl}=-d_{lk}=\frac{\langle k |p|l\rangle }{ -i \sqrt{\hbar
m\omega_{\rm pl}/2}}  .
     \end{equation}
For a weak nonlinearity ($N\gg 1$) one can derive (similar to Ref.\
\onlinecite{Martinis-bc}) the following approximations:
    \begin{eqnarray}
&& \frac{\epsilon_0}{\hbar \omega_{\rm pl}}=
\frac{1}{2}-\frac{11}{432 N}, \,\,\,
\frac{\epsilon_{k}-\epsilon_{k-1}}{\hbar \omega_{\rm pl}}= 1-\frac{5
 k}{36  N},
  \label{bc-appr1}\\
&&   c_{kk}=\frac{2k+1}{2\sqrt{3N}}, \,\,\,
    c_{01}\approx 1+\frac{0.0509}{N}, \,\,\,
c_{02} = \frac{-1}{\sqrt{54 N}},
     \qquad
    \label{bc-appr2}\\
&&    c_{03} = \frac{0.0170}{N}, \,\,\,  c_{12} =
\sqrt{2}+\frac{0.144}{N}, \qquad
    \label{bc-appr3}\\
&& d_{kk}=0, \,\,\, d_{01}=1 -\frac{0.0880}{N}, \,\,\, d_{02}=
-\frac{2}{\sqrt{54 N}}  ,
    \label{bc-appr4}\end{eqnarray}
where we keep only the lowest-order non-vanishing corrections, and
non-integer numbers are numerical results. In particular, these
approximations lead to Eq.\ (\ref{b,dw}) for
$b=(c_{11}-c_{00})\sqrt{\omega /\omega_{\rm pl}}$ and $\delta\omega
= (2\epsilon_1 -\epsilon_0-\epsilon_2)/\hbar$, where $\omega =
(\epsilon_1-\epsilon_0)/\hbar$. With next-order corrections, Eq.\
(\ref{b,dw}) becomes
  \begin{equation}
    b \approx \frac{1}{\sqrt{3N}}+\frac{0.28}{N^{3/2}} , \,\,\,
\frac{\delta \omega}{\omega} \approx \frac{5}{36N}+
\frac{0.18}{N^2}.
    \label{b,dw-2}\end{equation}

    Now let us discuss corrections for the analytics for $\Omega_{XX}$ and
$\Omega_{ZZ}$, which are next order in nonlinearity, while we still
use the lowest order in the perturbation theory. This modifies Eq.\
(\ref{Omega-XX-quant-lin1}) to become
    \begin{eqnarray}
    && \hspace{-0.5cm}   \Omega_{XX}= -\frac{2}{\hbar} K_{12}^a d_{01,1}d_{01,2} +
   \frac{2}{\hbar} c_{01,1} c_{01,2} \{K_{12}
    + K_{13}K_{23}
\nonumber \\
&& \hspace{-0.3cm} \times    [ \frac{ c_{01,3}^2}{E-\epsilon_{010}}
+ \frac{c_{01,3}^2}{E-\epsilon_{111}}
   + \frac{c_{02,3}^2}{E-\epsilon_{020}}
   + \frac{c_{02,3}^2}{E-\epsilon_{121}} ] \} , \qquad
   \label{Omega-XX-nonlin1}
    \end{eqnarray}
where the additional index $i=1,2,3$ in $c_{kl,i}$ and $d_{kl,i}$ is
the oscillator number. Now using Eqs.\
(\ref{bc-appr1})-(\ref{bc-appr4}) we obtain
        \begin{eqnarray}
&&    \Omega_{XX}=
   \frac{1+\frac{0.05}{N_1}+\frac{0.05}{N_2}}
    {\tilde L_4 \tilde L_5 \omega_{\rm qb} \sqrt{\tilde C_1 \tilde
    C_2}} \,  \left[ \tilde M- L_3^*
    \left( \frac{1+0.1/N_3}
    {1 -(\omega_{\rm qb}/\omega_3)^2}
 \right. \right.
\nonumber \\
&& \hspace{-0.5cm}  \left. \left. + \frac{0.01/N_3}
    {1-(\omega_{\rm qb}/2\omega_3)^2} ) \right) \right]
 +    \frac{(1-\frac{0.09}{N_1}-\frac{0.09}{N_2})\,
 \tilde C_a \sqrt{\tilde C_1 \tilde C_2}\, \omega_{\rm
qb}}{C_1 C_2} . \qquad
    \label{Omega-XX-nonlin2}\end{eqnarray}
    Equation (\ref{Omega-ZZ-quant1}) with account of next-order
corrections in nonlinearity becomes
    \begin{equation}
    \Omega_{ZZ} = \frac{b_1 b_2}{\hbar} \left[ K_{12} +
    \frac{2K_{13}K_{23} \, c_{01,3}^2}{\epsilon_{101}-\epsilon_{111}}+
    \frac{2K_{13}K_{23} \, c_{02,3}^2}{\epsilon_{101}-\epsilon_{121}}
    \right] ,
    \label{Omega-ZZ-quant4}\end{equation}
that modifies Eq.\ (\ref{Omega-ZZ-quant2}) [and the first term in
Eq.\ (\ref{Omega-ZZ-quant3})] to become
   \begin{equation}
   \Omega_{ZZ}=
   \frac{1+\frac{0.5}{N_1}+\frac{0.5}{N_2}}{6\sqrt{N_1N_2}}\,\,
    \frac{\tilde{M}-
L_{3}^* (1+0.1/N_3) }{\tilde{L}_4 \tilde{L}_5
\sqrt{\omega_1\omega_2} \sqrt{C_1C_2}} .
    \label{Omega-ZZ-quant5}
    \end{equation}
A useful technical trick in deriving these results [as well as Eq.\
(\ref{Omega-ZZ-quant1})] is to slightly shift the coordinates
$\delta\phi_{i}$ for the three oscillators (by slightly changing
$\{\phi_{1,\rm st}, \phi_{2,\rm st}, \phi_{3,\rm st} \}$) to produce
$\langle 0|\delta \phi_{i}|0\rangle =0$ for each oscillator; this
significantly reduces the number of terms due to nonlinearity in the
derivation.



\begin{thebibliography}{99}
\bibitem{magnetostatics-sign}
The classical magnetic  energy in the coupler is $E_m=L_4 I_4^2/2
+L_5 I_5^2/2-MI_4 I_5$, where $I_4$ and $I_5$ are the currents
through the inductors $L_4$ and $L_5$ (flowing in the ``down''
direction in Fig.\ 1). When expressed in terms of the fluxes
$\Phi_4=L_4 I_4-M I_5$ and $\Phi_5=L_5 I_5 -M I_4$, it becomes
$E_m=\Phi_4^2/2KL_4+\Phi_5^2/2KL_5+(M/KL_4L_5)\Phi_4\Phi_5$, where
$K=1-M^2/L_4L_5$. Notice the sign change of the interaction term,
which is important in Eqs.\ (\protect\ref{Ham1}) and
(\protect\ref{Omega-ZZ-class}).
\bibitem{exp-dif} In the experiment \protect\cite{Bialczak} the
regime of the OFF coupling was actually realized at the bias current
$I_B$ close to zero. This happened because of designed $M/L_3\approx
1$ at $I_B=0$. We think that a better mode is when the coupling is
OFF at $|I_B|/I_{3,\rm cr}\alt 1$, and consider $M/L_3\simeq 2$ at
$I_B=0$.
\end{thebibliography}
\end{document}